\documentclass{aa}

\usepackage{graphicx}
\usepackage{ulem}
\usepackage{txfonts}

\usepackage{amsmath,amssymb}
\usepackage{epsfig}
\usepackage{morefloats}
\usepackage{rotating}
\usepackage{soul}

\def\beq{\begin{equation}}
\def\eeq{\end{equation}}
\def\beqa{\begin{eqnarray}}
\def\eeqa{\end{eqnarray}}
\def\ban{\begin{eqnarray*}}
\def\ean{\end{eqnarray*}}
\def\bi{\begin{itemize}}
\def\ei{\end{itemize}}

\newcommand{\pc}[1]{\ensuremath{\left(#1\right)}}

\begin{document}

\title{Influence of the tetraneutron on the EoS under \\ core-collapse supernovae and heavy-ion collisions conditions}

\author{Helena Pais\inst{1} \and Conrado Albertus\inst{1} \and M. \'Angeles P\'erez-Garc\'ia\inst{1} \and Constan\c ca Provid\^encia\inst{2}}

\institute{Department of Fundamental Physics, University of Salamanca, 37008 Salamanca, Spain.  \and
CFisUC, University of Coimbra, P-3004-516 Coimbra, Portugal. } 

\date{Received xxx Accepted xxx}

 \abstract
   {Recently, a resonant state of four neutrons (tetraneutron) with an energy of $E_{4n}=2.37\pm 0.38 \rm{(stat)} \pm 0.44 \rm{(sys)}$ MeV and a width of $\Gamma=1.75\pm 0.22 \rm{(stat)} \pm 0.30 \rm{(sys)}$ MeV was reported.}
   {In this work, we analyse the effect of including such an exotic state on the yields of other light clusters, that not only form in astrophysical sites, such as core-collapse supernovae and neutron star mergers, but also in heavy-ion collisions.}
   {To this aim, we use a relativistic mean-field formalism, where we consider in-medium effects in a two-fold way, via the couplings  of the clusters to the mesons, and via a binding energy shift, to compute the low-density equation of state for nuclear matter at finite temperature and fixed proton fraction. We consider five light clusters, deuterons, tritons, heliums, $\alpha$-particles, and $^6$He, immersed in a gas of protons and neutrons, and we calculate their abundances and chemical equilibrium constants with and without the tetraneutron. We also analyse how the associated energy of the tetraneutron would influence such results.}
   {We find that the low-temperature, neutron-rich systems, are the ones most affected by the presence of the tetraneutron, making neutron stars excellent environments for their formation. Moreover, its presence in strongly asymmetric matter may increase considerably the proton and the $\alpha$-particle fractions. This may have an influence on the dissolution of the accretion disk of the merger of two neutron stars.}
   {}


\authorrunning{Pais et al.}

\maketitle

\section{\label{intro} Introduction}

For decades the experimentalists have been trying to find a resonant/bound state constituted by four neutrons. Recently in an experiment at RIKEN with the SAMURAI spectrometer, using a high-energy beam of $^8$He on a proton target, \cite{Duer22} reported on the production of a resonant state of four neutrons with an energy of $E_{4n}=2.37\pm 0.38 \rm{(stat)} \pm 0.44 \rm{(sys)}$ MeV and a width of $\Gamma=1.75\pm 0.22 \rm{(stat)} \pm 0.30 \rm{(sys)}$ MeV. This value for the energy is substantially higher (and the width lower) than a previous experiment \citep{Kisamori2016}, where a high-energy radioactive $^8$He beam was hit on a liquid $^4$He target, producing a resonant tetraneutron state, with an energy of $E_{4n}=0.8 \pm 1.4$ MeV, with $\Gamma=2.6$MeV as an upper limit, though the uncertainty reported was quite large. The first report, of a possible bound state was given in 2002 by \cite{Marques2002}, from a a reaction $^{14}$Be$\rightarrow ^{10}$Be$+4n$, and later in 2005 \citep{Marques2005}, where instead a resonant state with an energy of $\sim 2$MeV, was reported, the former result being consistent with this value.

Neutron stars (NS), and other astrophysical phenomena, such as core-collapse supernovae or neutron star  mergers, are systems where this exotic tetraneutron state may naturally be present, see \cite{Ivanytskyi19}. These latter systems are characterized by finite temperatures, that may reach 50 to 100 MeV \citep{Oertel2017}, do not attain $\beta-$equilibrium, like NS, and fixed proton fractions must be considered. They also gather ideal conditions for the formation of light clusters, such as $\alpha-$particles, tritons, deuterons or $^3$He, that may in turn affect the cooling of these objects,  as the neutrino mean-free path may be changed \citep{Arcones2008}. This may then have an impact on the transport properties and dynamics of these astrophysical systems. Moreover, in the merger of two neutron stars, these clusters may also have an influence on the dissolution of the accretion disk \citep{Rosswog2015}, and on the fraction of the ejected material \citep{Bauswein2013} that determines the features of their associated Kilonovae light curves \citep{Perezetal2022AA, Prada}. 

These clusters are not only formed in protoNS or NS mergers, but are also produced in heavy-ion collisions (HIC) on Earth. In 2012, \cite{Qin2012} reported a finite-temperature constraint on cluster yields, from the analysis of the production of four light clusters at the NIMROD detector. Later, the INDRA detector was able to measure the yields of five light clusters \citep{Bougault2020}, and an analysis, in which the systems in question were not treated under an ideal-gas assumption, was performed \citep{PaisPRL,PaisJPG}.

From a theoretical point view, there are several formalisms one might consider to describe  the sub-saturation EoS with light clusters, from the single-nucleus approximation, like the Lattimer and Swesty EoS \citep{LS91}, or the nuclear statistical equilibrium models such as e.g. \cite{Raduta10,Hempel10}, that consider all possible nuclear clusters in equilibrium, or even a quantum statistical approach \citep{Roepke15}, that also takes into account the excited states.
In this work, we use a relativistic mean-field formalism (see e.g. \citep{Pais18} and references therein), where the clusters are considered as new degrees of freedom, and have an effective mass that depends on the density. They are described as point-like, and in-medium effects are taken into account in a two-fold way: via a binding energy shift, constructed \textit{à la} Thomas-Fermi, which works as an excluded-volume effect, that not only prevents double-counting of the clusters but also enforces their dissolution, and also via a factor in the scalar cluster-meson couplings, that was fitted \citep{Pais18} to the Virial EoS \citep{virial1,virial2,virial3}, and later to the INDRA data \citep{PaisPRL}. The formalism used for in-medium modifications on the EoS was derived in \cite{Pais18}. Here we adapt it to include the tetraneutron, just as in the same spirit of \cite{Pais2019}, where light clusters with mass number $A$ up to 12 were considered. We do not include a geometrical excluded-volume prescription, such as in a NSE framework, see e.g. \cite{Hempel10,Raduta19}. We chose the FSU EoS \citep{FSU}, that, even though it does not produce two-solar-mass stars, describes well the sub-saturation range of the EoS, since it was fitted to reproduce the properties of nuclear matter properties at saturation. 

In order to understand how the presence of the $4n$ resonance can affect the properties of hot nuclear matter, in particular as concerns composition, we perform a study of hot non-homogeneous matter, considering two possible proton fractions, typical of astrophysical systems, such as core-collapse supernova matter or binary neutron star mergers. There is a previous study on the appearance of tetraneutrons at low temperature \citep{Ivanytskyi19}, where the authors consider this exotic cluster and find that the tetraneutron presence in neutron rich matter could significantly impact the nucleon pairing fractions and the distribution of baryonic charge among species. They consider different conditions i.e. zero temperature and $\beta-$equilibrium matter using a different formalism including explicit treatment of Bose-Einstein condensates of tetraneutrons. 
In that work, the interaction of the $4n$ resonance introduced an excluded volume within the Van der Waals approximation besides the vector meson driven repulsion and Pauli blocking for a model which did not include the $^6$He cluster compared to the present study. In the present approach all species are treated as point-like so that an  excluded-volume mechanism is avoided. The vector meson and binding energy shifts take into account finite size and Pauli blocking effects, and the model parameters are fitted to HIC data.   Besides, the temperatures considered are above the condensation critical temperature. 

We consider a system of 5 light clusters, $^2$H, $^3$H, $^3$He, $^4$He, $^6$He, i.e. all five clusters measured by INDRA, and we also include the tetraneutron, $4n$, immersed in a gas of protons and neutrons. The inclusion of the tetraneutron will follow  the data available from recent observations of this cluster as a resonant state,  as measured in \cite{Duer22}. We will consider $E_{4n}$ as having the average value of -2.37 MeV, implying that we consider the cluster as a resonant state, but we will start by discussing how the uncertainty on this quantity may affect the results. We will then address the effect of the $4n$  in the mass fractions of the other clusters, taking into account several possible temperatures, 4, 10 and 20 MeV, and proton fractions equal to 0.1 and 0.3 as found in $\beta$-equilibrium mater and in core-collapse supernova matter. It will be shown that the $4n$ resonance may have an important effect on the clusters abundances, in particular, for very asymmetric and not too hot matter. If the model fitted to the INDRA data is considered, these effects may imply an increase of a factor of 4 to 5 on the proton and $^4$He abundances, respectively, for $T=4$ MeV and a density of $\sim 0.02$ fm$^{-3}$.

The structure of the article is as follows: in Section II, the RMF formalism is briefly described, then in the next Section, the results are presented, and, finally, in the last Section, some conclusions are drawn.

\section{Formalism}

 \begin{figure*}
 \begin{tabular}{cc}
 \includegraphics[width=0.45\textwidth]{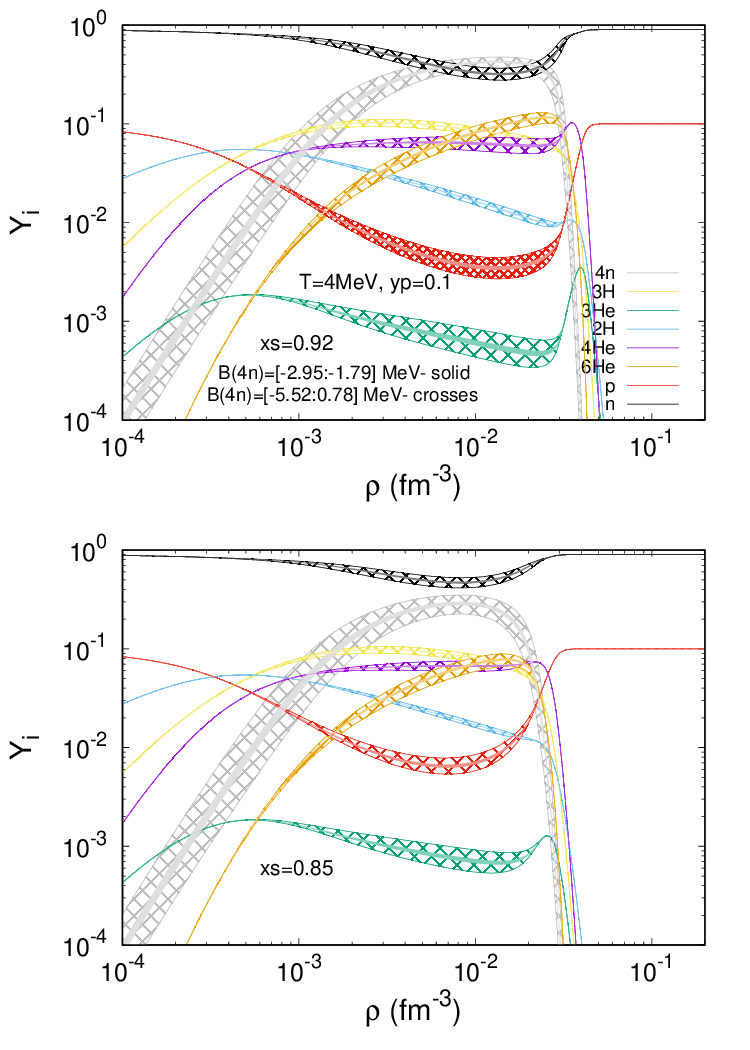} & \includegraphics[width=0.45\textwidth]{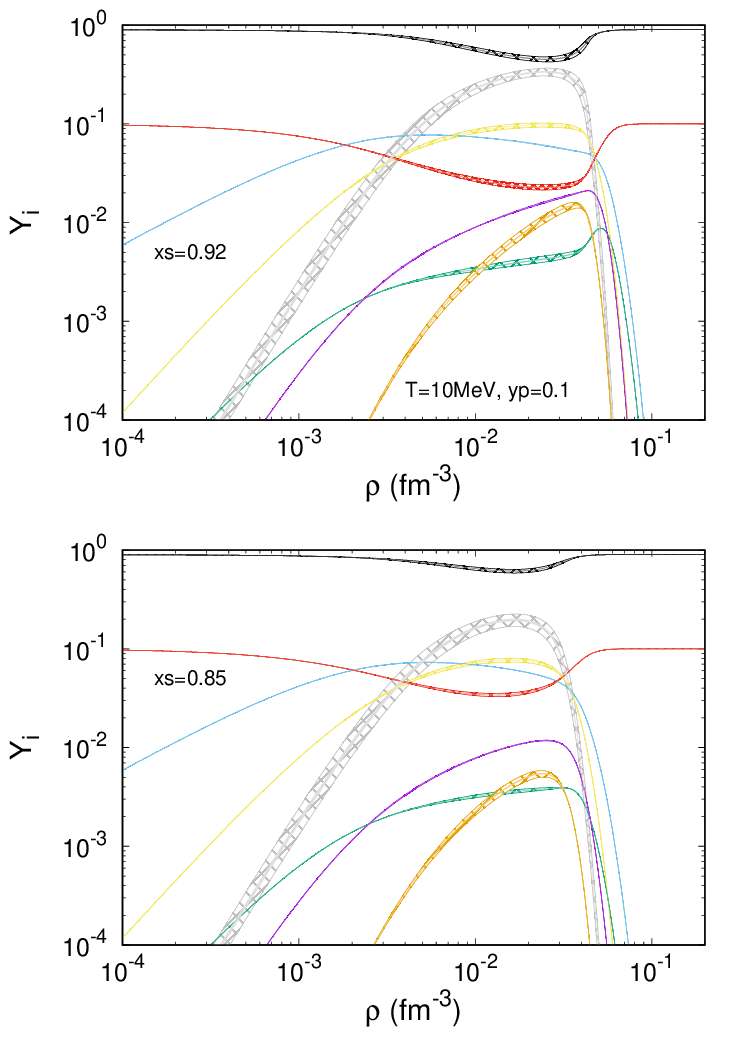}
 \end{tabular}
  \caption{The mass fractions as a function of the density, when considering different energy bands for the tetraneutron, for $T=4$ (left) and 10 (right) MeV, and two different scalar couplings for the clusters, $x_s=0.92$ (top) and $x_s=0.85$ (bottom).}
\label{fig0}
\end{figure*}

Our system includes light clusters, both bosons, deuterons ($^2$H), $\alpha$-particles ($^4$He), $^6$He and the tetraneutron $4n$,
and fermions, tritons ($^3$H) and helions ($^3$He), immersed in a gas of  neutrons ($n$) and protons ($p$).

The Lagrangian density is given by \citep{typel10,ferreira12,pais15,avancini17,Pais18,Pais2019}:
\begin{eqnarray}
{\cal L} &=& \sum_{j=n,p,d,t,h,\alpha,^6{\rm He},4n} {\cal L}_{j}                            
+ {\cal L}_{\sigma} + {\cal L}_{\omega} + {\cal L}_{\rho}+ {\cal L}_{\omega\rho}.
\end{eqnarray}
The couplings of the clusters to the mesons are
defined in terms of the couplings of the nucleons, $g_s,\, g_v,\, g_\rho$ to the $\sigma$, $\omega$ and
$\rho$-mesons, respectively. 
For the fermionic clusters,  $j=t,h$, we have:
\begin{eqnarray}
{\cal L}_j &=& \bar{\psi}\left[\gamma_\mu i D_j^\mu - M_j^*\right]\psi,
\end{eqnarray}
with  
\begin{equation}
iD^{\mu }_j = i \partial ^{\mu }-g_{vj} \omega^{\mu }-\frac{g_\rho}{2}{\boldsymbol\tau}_j \cdot \mathbf{b}^\mu ,
\end{equation}
where ${\boldsymbol \tau}_j$ are the Pauli matrices and $g_{vj}$ is the
  coupling of cluster $j$ to the vector meson $\omega$ and, it is defined
  as $g_{vj}=A_j g_v$ for all clusters.

The Lagrangian density for the bosonic clusters, $j=d,\alpha,^6{\rm He},4n$, is given by
\begin{eqnarray}
\mathcal{L}_{k=\alpha,4n }&=&\frac{1}{2} (i D^{\mu}_{k} \phi_{k})^*
(i D_{\mu k} \phi_{k})-\frac{1}{2}\phi_{k}^* \pc{M_{k}^*}^2
\phi_{k},\\
\mathcal{L}_{d}&=&\frac{1}{4} (i D^{\mu}_{d} \phi^{\nu}_{d}-
i D^{\nu}_{d} \phi^{\mu}_{d})^*
(i D_{d\mu} \phi_{d\nu}-i D_{d\nu} \phi_{d\mu})\nonumber\\
&&-\frac{1}{2}\phi^{\mu *}_{d} \pc{M_{d}^*}^2 \phi_{d\mu},
\end{eqnarray}
with
\begin{equation}
iD^{\mu }_j = i \partial ^{\mu }-g_{vj} \omega^{\mu }.
\end{equation}

For the nucleonic gas,  $j=n,p$, we have:
\begin{eqnarray}
{\cal L}_j &=& \bar{\psi}\left[\gamma_\mu i D^\mu - m^*\right]\psi
\end{eqnarray}
with
\begin{eqnarray}
i D^\mu&=&i\partial^\mu-g_v\omega^\mu-\frac{g_\rho}{2}{\boldsymbol\tau}_j \cdot \mathbf{b}^\mu \\
m^*&=&m-g_s\phi_0 \label{meff}
\end{eqnarray}

For the fields, we have the standard RMF expressions:
\begin{eqnarray}
{\cal L}_\sigma&=&+\frac{1}{2}\left(\partial_{\mu}\phi\partial^{\mu}\phi
-m_s^2 \phi^2 - \frac{1}{3}\kappa \phi^3 -\frac{1}{12}\lambda\phi^4\right),\nonumber\\
{\cal L}_\omega&=&-\frac{1}{4}\Omega_{\mu\nu}\Omega^{\mu\nu}+\frac{1}{2}
m_v^2 V_{\mu}V^{\mu}, \nonumber \\ 
{\cal L}_\rho&=&-\frac{1}{4}\mathbf B_{\mu\nu}\cdot\mathbf B^{\mu\nu}+\frac{1}{2}
m_\rho^2 \mathbf b_{\mu}\cdot \mathbf b^{\mu}, \nonumber \\ 
{\cal L}_{\omega\rho}&=& g_{\omega\rho} g_\rho^2 g_v^2 V_{\mu}V^{\mu}\mathbf b_{\nu}\cdot \mathbf b^{\nu}
\end{eqnarray}
where
$\Omega_{\mu\nu}=\partial_{\mu}V_{\nu}-\partial_{\nu}V_{\mu}, $ and $ \mathbf B_{\mu\nu}=\partial_{\mu}\mathbf b_{\nu}-\partial_{\nu} \mathbf b_{\mu}
- g_\rho (\mathbf b_\mu \times \mathbf b_\nu)$.

\begin{figure*}
 \begin{tabular}{ccc}
 \includegraphics[width=0.98\textwidth]{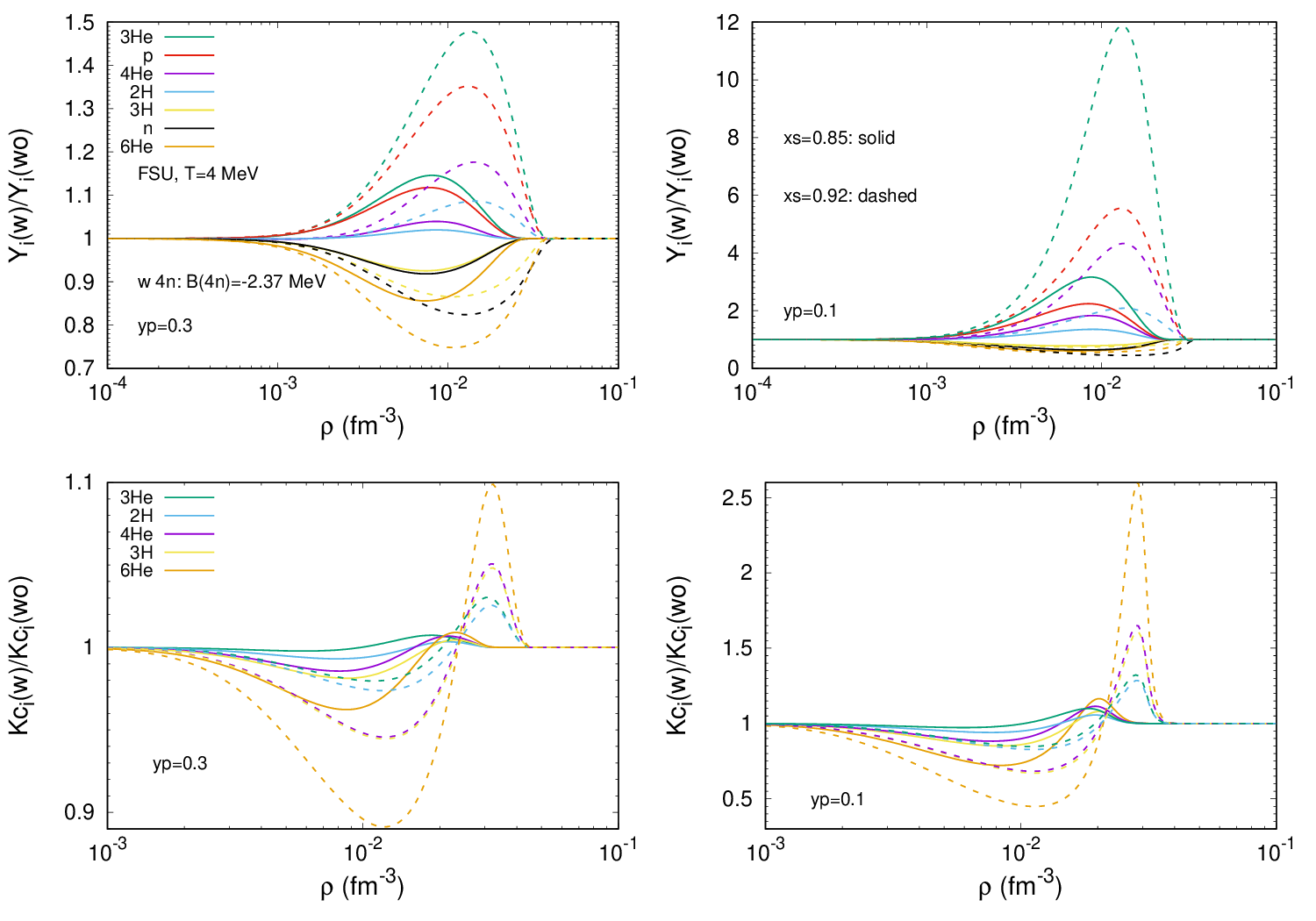}
 \end{tabular}
  \caption{(Top) Ratio of the  mass fractions of the clusters and of the gas with ($Y_i(w)$) and without ($Y_i(wo)$) the tetraneutron, for a fixed proton fraction of $y_p=0.3$ (left), and $y_p=0.1$ (right), in a calculation where we consider a fixed binding energy for the $4n$, taken as $B_0(4n)=-2.37$ MeV. (Bottom) The chemical equilibrium constants, $K_c[i]$, in a calculation with ($w$) and without ($wo$) the tetraenutron, in the same conditions as the top panels. In all panels, the temperature is fixed to 4 MeV, the FSU EoS is considered, and the scalar cluster-meson coupling is taken  as $x_s=0.85$ (solid) and $x_s=0.92$ (dashed). }
\label{fig1}
\end{figure*}

The total binding energy of a light cluster $j$ is given by 
\begin{eqnarray}
B_j=A_j m^*-M_j^* \,, \quad j=d,t,h,\alpha,^6{\rm He},4n \,, \label{binding}
\end{eqnarray}
with $M_j^*$ the effective mass of cluster $j$, which is determined by
the meson coupling  as well as by a binding energy shift: 
\begin{eqnarray}
M_j^*=A_j m - g_{sj}\phi_0 - \left(B_j^0 + \delta B_j\right) \, .
\label{meffi2}
\end{eqnarray}

In expression (\ref{meffi2}), $B^0_j$ is the binding energy of the cluster $j=^2$H,$^3$H,$^3$He,$^4$He,$^6$He in the vacuum, and these constants are fixed to experimental values. For the tetraneutron, we take the values of \cite{Duer22}. We will take the binding energy of the tetraneutron, $B_{4n}^0$, as negative, since it was considered a resonant state in \cite{Duer22}, while for the other 5 light clusters, being bound states, they will have their binding energies  positive.

The binding energy shift  $\delta B_j$ is given by \citep{Pais18}
\begin{eqnarray}
\delta
  B_j=\frac{Z_j}{\rho_0}\left(\epsilon_p^*-m\rho_p^*\right)+\frac{N_j}{\rho_0}\left(\epsilon_n^*-m\rho_n^*\right)
  \, .
\label{deltaB}
\end{eqnarray}
 This term acts as the energetic counterpart of the excluded volume mechanism in the Thomas-Fermi approximation. $\rho_0$ is the nuclear saturation density, and $\epsilon_j^*$ and $\rho_j^*$  are the energy density and density of the gas lowest energy levels, respectively. This means that the energy states occupied by the gas are removed from the calculation of the cluster binding energy, thus avoiding double counting of the particles of the gas and the ones of the clusters.

Regarding the scalar and vector cluster-meson couplings, we follow the prescription introduced in \cite{Pais18}. The scalar coupling is given by
\begin{equation}
\label{gs}
g_{sj}=x_{sj} A_j g_s, 
\end{equation}
 while the vector coupling is given by
\begin{equation}
\label{gv}
g_{vj}=A_j g_v.
\end{equation}

$A_j$ corresponds to the number of nucleons in cluster $j$. The $x_s$ factor can vary from 0 to 1. In a previous work, its value was fixed to $x_s=0.85\pm0.05$ from a fit to the Virial EoS. In later works, the value was found to be higher, $x_s=0.92\pm0.02$, when a fit to experimental data was considered \citep{PaisPRL,PaisJPG}. In the following, we will use both couplings to test its effect in the clusters abundances. The dissolution of the clusters will be affected by a combination of both the binding energy shift, $\delta B_j$, and this factor $x_s$. Substituting eqs.~(\ref{meffi2}), (\ref{meff}) and (\ref{gs}) in eq.~(\ref{binding}), we obtain
\begin{eqnarray}
B_j&=&A_j g_s \phi_0 \left(x_{sj}-1 \right) + B_j^0 + \delta B_j \, .
\end{eqnarray}
 For the two extreme cases, we have
\begin{eqnarray}
B_j&=&B_j^0 + \delta B_j \, , {\rm if} \,  x_{sj}=1 \, , \\
B_j&=&B_j^0 + \delta B_j  - A_j g_s \phi_0  \, , {\rm if} \,  x_{sj}=0 \, .
\end{eqnarray}

This implies that a larger $x_{sj}$ corresponds to a larger binding energy,  and, consequently, the dissolution of the cluster will occur at larger densities.  If $x_{sj}=1$, the dissolution is totally defined by the binding shift $\delta B_j$. Notice that at finite  temperature, the clusters  dissolve at a density well above the one for which $B_j\sim 0$. For this reason the tetraneutron survives even as a resonance. The larger the temperature the more the fraction of clusters is defined by their mass and isospin, and not by the binding energy.

With the same set of couplings determined in the last section, we
calculate the chemical equilibrium constants 
\begin{equation}
K_c[j]=\frac{\rho_j}{\rho_n^{N_j}\rho_p^{Z_j}}
\label{kc}
\end{equation}
where $\rho_j$ is the number density of cluster $j$, with neutron
number $N_j$  and  proton number $Z_j$, and $\rho_p$, $\rho_n$ are,
respectively, the number densities of free protons and neutrons.

Even though there are no experimental $K_c$ for the tetraneutron, we will calculate it for the other clusters, considering calculations where we do and do not include the $4n$. This may give a hint on the abundance of the clusters, and the presence or not of the tetraneutron.

Let us also refer to another point that must be discussed. The tetraneutron, just as the other light clusters in this work, is treated as a point-like particle, and one may ask whether or not an exclusion volume should be included in order that the model does not break down as the temperature increases. In this model, the role of the exclusion volume included, for instance, in \cite{LS91} or in \cite{Shen1998}, is undertaken by the $\omega$-meson, as clearly discussed in \cite{typel10} and \cite{Avancini2010}. We, therefore, will not consider any explicit  exclusion volume term.

\section{Results}
\label{sec:results}

\begin{figure*}
 \begin{tabular}{ccc}
 \includegraphics[width=0.98\textwidth]{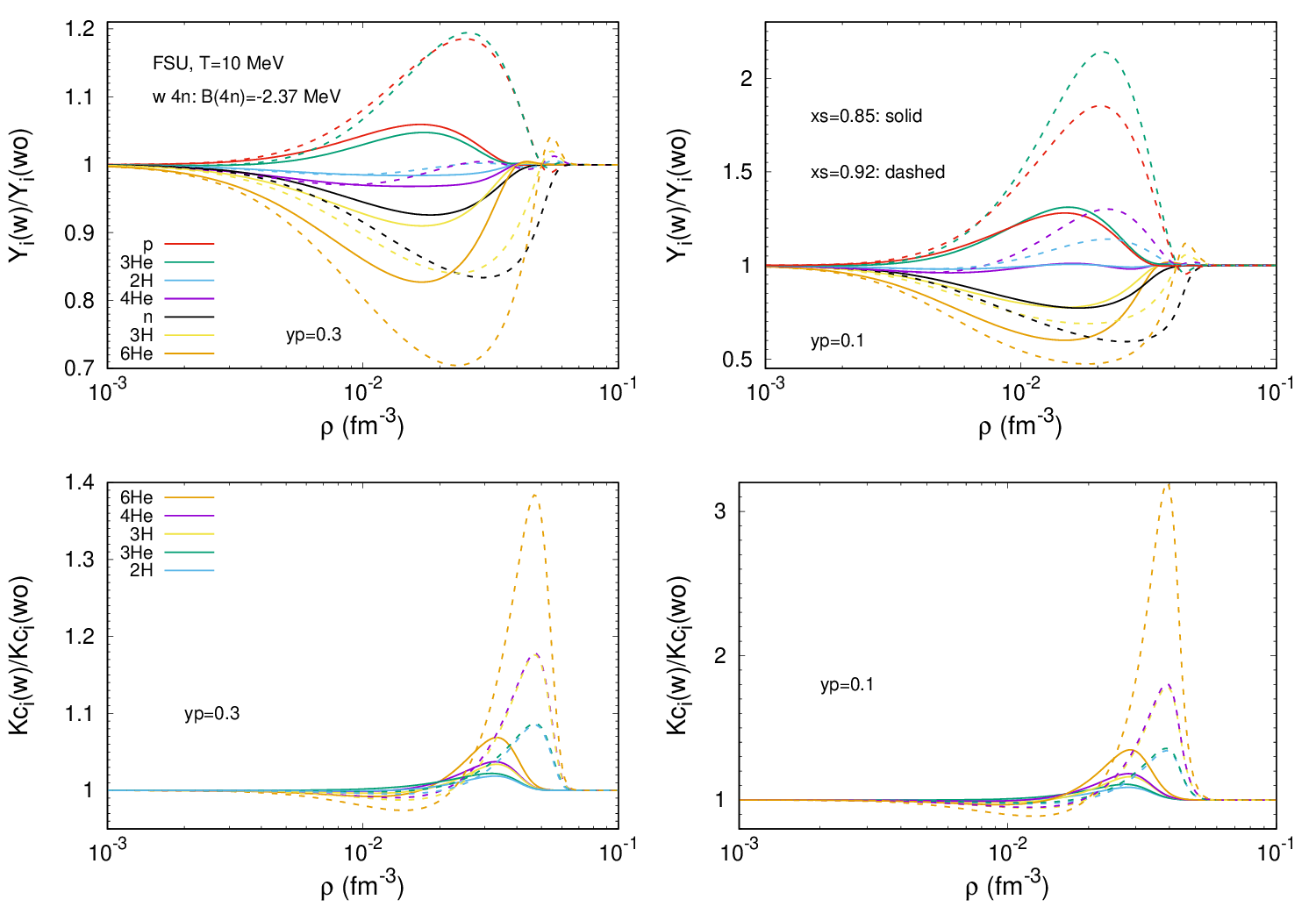}
 \end{tabular}
  \caption{(Top) Ratio of the  mass fractions of the clusters and of the gas with ($Y_i(w)$) and without ($Y_i(wo)$) the tetraneutron, for a fixed proton fraction of $y_p=0.3$ (left), and $y_p=0.1$ (right), in a calculation where we consider a fixed binding energy for the $4n$, taken as $B_0(4n)=-2.37$ MeV. (Bottom) The chemical equilibrium constants, $K_c[i]$, in a calculation with ($w$) and without ($wo$) the tetraenutron, in the same conditions as the top panels. In all panels, the temperature is fixed to 10 MeV, the FSU EoS is considered, and the scalar cluster-meson coupling is taken  as $x_s=0.85$ (solid) and $x_s=0.92$ (dashed). Notice the different y-axis scales.}
\label{fig2}
\end{figure*}

\begin{figure*}
 \begin{tabular}{ccc}
 \includegraphics[width=0.98\textwidth]{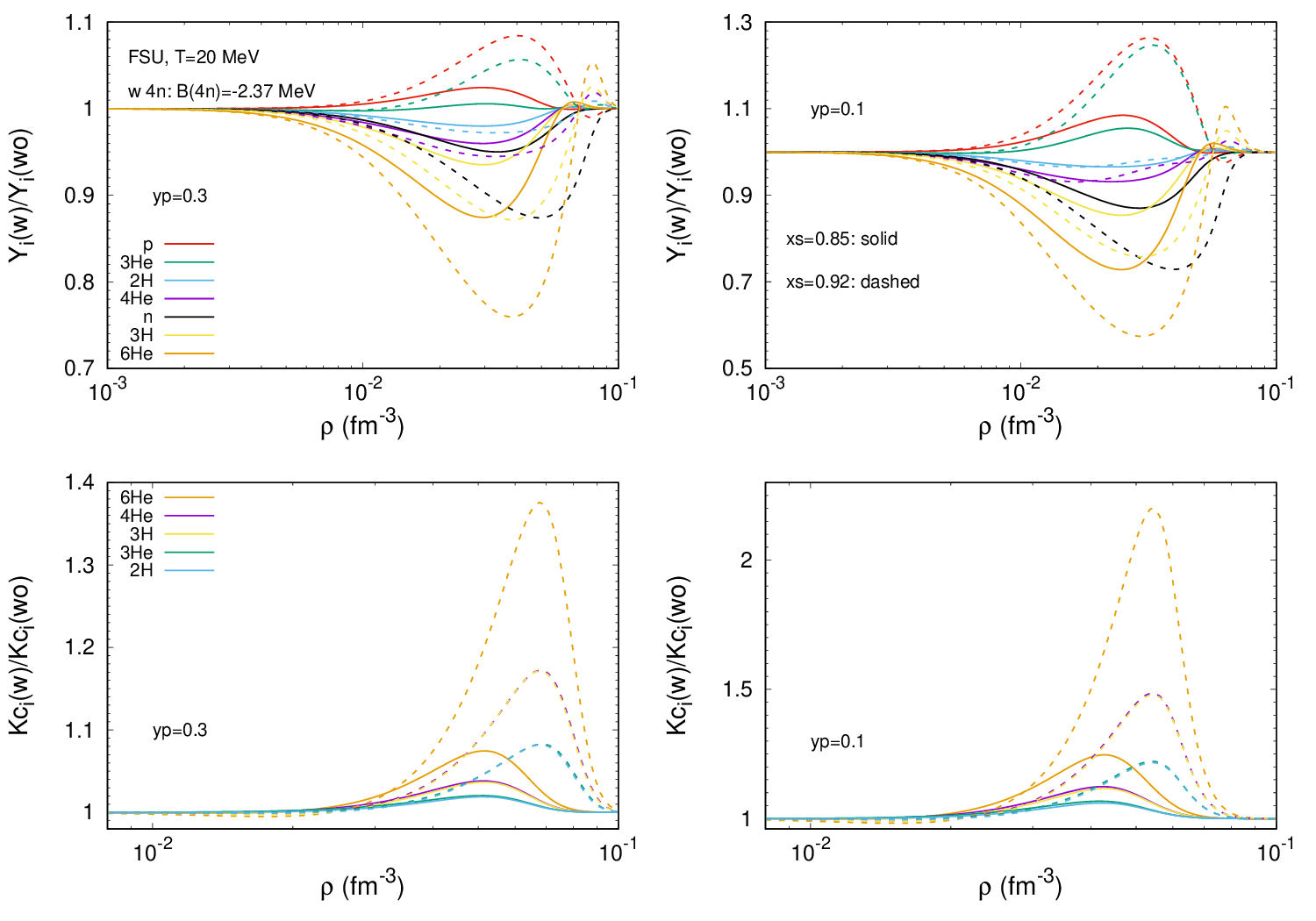}
 \end{tabular}
  \caption{(Top) Ratio of the  mass fractions of the clusters and of the gas with ($Y_i(w)$) and without ($Y_i(wo)$) the tetraneutron, for a fixed proton fraction of $y_p=0.3$ (left), and $y_p=0.1$ (right), in a calculation where we consider a fixed binding energy for the $4n$, taken as $B_0(4n)=-2.37$ MeV. (Bottom) The chemical equilibrium constants, $K_c[i]$, in a calculation with ($w$) and without ($wo$) the tetraenutron, in the same conditions as the top panels. In all panels, the temperature is fixed to 20 MeV, the FSU EoS is considered, and the scalar cluster-meson coupling is taken  as $x_s=0.85$ (solid) and $x_s=0.92$ (dashed). }
\label{fig3}
\end{figure*}

\begin{figure*}
 \begin{tabular}{ccc}
 \includegraphics[width=0.98\textwidth]{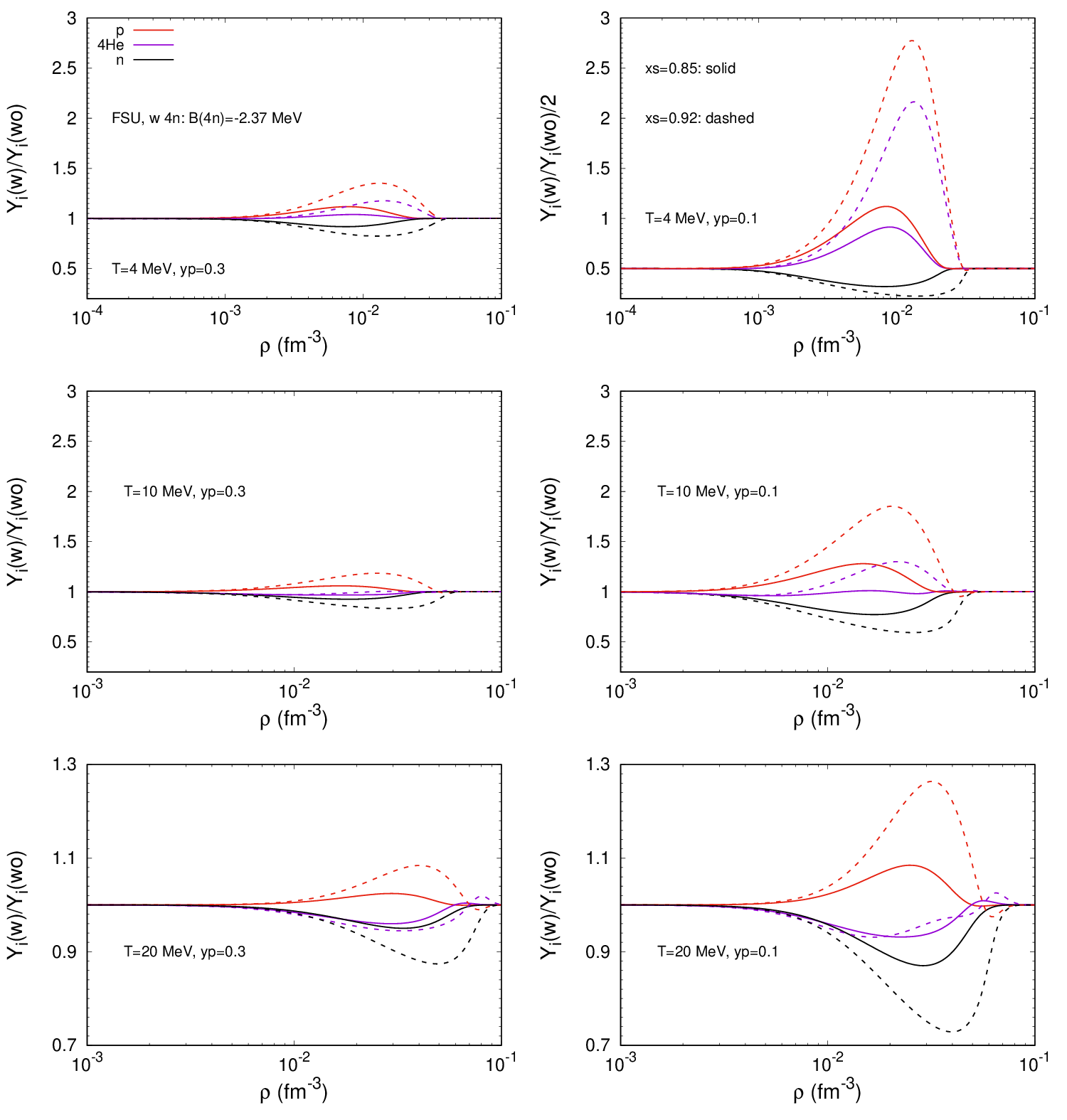}
 \end{tabular}
  \caption{(Color online) The ratio of the mass fractions of the proton and neutron gas  and of the $\alpha$ particles with ($Y_i(w)$) and without ($Y_i(wo)$) the tetraneutron for a fixed proton fraction of $y_p=0.3$ (left) and $y_p=0.1$ (right) and $T=4$ (top), 10 (middle) and 20 (bottom) MeV, in a calculation where we take the binding energy of the $4n$ as -2.37 MeV. In all panels, the FSU EoS is considered,  and the scalar cluster-meson coupling is taken  as $x_s=0.85$ (solid) and $x_s=0.92$ (dashed). Notice that the y-axis scale of the bottom panels is different, and the ratio of the top right panel is divided by 2.}
\label{fig4}
\end{figure*}

In this section, we show the mass fractions of the clusters and discuss how they are affected by temperature and isospin asymmetry of the medium. We define them as
\begin{eqnarray}
Y_i&=& A_i \frac{\rho_i}{\rho_B} \, .
\end{eqnarray}

In particular, we will discuss how the fraction of the classical clusters are affected by the presence of the tetraneutron. For the couplings $x_{sj}$, we will take  the values  0.85 and 0.92 as determined respectively in \cite{Pais18}  and \cite{PaisPRL}, fitting our model to two different sets of experimental data \citep{Qin2012,Bougault2020}. The value $x_{sj}=0.85$ was found from a fit to the Virial EoS \citep{Pais18}. Later, when fitting other experimental data, from the INDRA collaboration, we found that the calibration would need a larger coupling, which means larger binding energies, and therefore a larger dissolution density. This value was found to be $x_{sj}=0.92$ \citep{PaisPRL}.

\subsection{On the energy of the tetraneutron}

Here we show the different abundances one would get when considering different energies for the tetraneutron.

We will consider its energy given by two bands, with the following extremes:
\begin{eqnarray}
B^0_{4n}&=& -2.37 \pm \sqrt{0.38^2 + 0.44^2} = [-2.95:-1.79] \quad \label{b04n1} \\    
B^0_{4n}&=& -2.37 \pm 1.8\Gamma = [-5.52:0.78] \,     \label{b04n2}
\end{eqnarray}
both defined in MeV. The width $\Gamma$ is equal 1.75 MeV.
Eq.~(\ref{b04n1}) considers both the systematic and statistical errors in the energy value, and does not take into account the width of the resonance, whereas eq.~(\ref{b04n2}) considers $\Gamma$, multiplied by a factor, obtained if the width distribution is Cauchy-type. The derivation of this factor is given in Appendix~\ref{append}. Note that the tetraneutron properties are yet uncertain and have varied from previous calculations \citep{Ivanytskyi19}.

Taking these binding energies and uncertainties, we calculate and discuss how the $4n$ and light cluster abundances vary with the density.  In Fig.~\ref{fig0}, the mass fractions of the clusters $d,\, t,\, h, \, \alpha$, $^6$He and $4n$ are plotted for a proton fraction typical of NS, $y_p=0.1$, and two representative temperatures, 4 and 10 MeV. Besides we also consider two values of the coupling of the clusters to the $\sigma$-meson, as explained above. 
In this Figure, we take for the $4n$ the binding energy defined by the range given in Eq.~(\ref{b04n1}), given by the solid bands. The crossed bands are obtained taking, instead, the range defined by Eq.~(\ref{b04n2}). These define slightly wider regions, but the same overall behavior is obtained.

For this small proton fraction, it is striking that at the maximum of the clusters, the mass fraction of the $4n$ cluster is the largest among the clusters and can be as large as the one of free neutrons if $x_{sj}=0.92$, and just slightly smaller for $x_{sj}=0.85$. On the other hand, $4n$ is the first cluster to dissolve. At $T=10$ MeV, the $4n$ is still the most abundant cluster but the impact on the free nucleons is not so strong. For a temperature of 10 MeV, the neutron content and the  magnitude of the mass (and not the binding energy) is defining the cluster abundances: $4n$ are still the most abundant at $\rho\sim0.02$fm$^{-3}$ due to the neutron content, next comes  the $d$ and the $h$, the first one because it is the lightest cluster and the other because it is the  next cluster in mass with the largest neutron content.

We see that the abundance of the tetraneutron increases for the low-$T$ and high $x_{sj}$ case. We also observe that by even considering two different ranges for the energy of the $4n$, the difference in the abundance of the other clusters is not much. That difference is only non-negligible at the maximum of the $4n$ abundance, i.e. both at the onset and dissolution, the difference is completely negligible. 

This implies that such a binding energy of the tetraneutron won't make much difference in the systems considered, typical of core-collapse supernovae or heavy-ion collisions, where these clusters are also measured. The impact of such cluster would be higher in more neutron-rich systems.

In order to better quantify the effect of the inclusion of the $4n$ clusters, in the next section we concentrate our discussion on the ratios of quantities calculated with and without the tetraneutron.

\subsection{Effect of including $4n$}

In this subsection we  compare the effect of including $4n$ in the matter, by defining particle fractions and equilibrium constant ratios between the quantities obtained with and without the inclusion of $4n$ particles.

In Figs.~\ref{fig1}, \ref{fig2} and \ref{fig3}, we plot the ratio of the mass fractions of the five light clusters and of the gas, and of the chemical equilibrium constants of the five light clusters, in a calculation with and without the tetraneutron against the baryonic density, using the FSU model, a fixed temperature of 4, 10 and 20 MeV and two proton fractions, $y_p=0.1$, and 0.3. The scalar cluster-meson coupling is chosen as $x_{sj}=0.85$ and 0.92. The energy of the tetraneutron is chosen as $B^0_{4n}=-2.37$ MeV, which is the average value of \cite{Duer22}, taken as a negative value to consider it as a resonant state, as opposed to a bound state, like the other clusters considered, whose binding energy value in the vacuum is taken to be positive. We observe that all clusters dissolve, including the tetraneutron, below 0.1 fm$^{-3}$. The fraction maxima move from $\sim 0.01$fm$^{-3}$ at $T=4$ MeV to $\sim 0.05$fm$^{-3}$ for $T=20$MeV.

The role of the scalar cluster-meson coupling in the clusters' abundances should also be discussed.  Considering both values allows us to see how it reflects on the mass fractions, and estimate the uncertainty connected to the calibration of the underlying model used to perform the calculations.

For the lowest temperature considered, shown in Fig.~\ref{fig1}, we observe that the proton-rich and symmetric clusters increase their abundance, such as $^3$He or $^4$He, whereas the neutron-rich cluster yields decrease, as the neutrons are being consumed by the tetraneutron.  The effect is larger for the low-temperature and low-proton fraction system, {being in line with previous zero-temperature calculations \citep{Ivanytskyi19}}. When we increase the temperature, i.e., in Figs.~\ref{fig2} and \ref{fig3}, we see that the abundance of the proton-rich clusters becomes smaller, and looking e.g. at the $\alpha-$particle, for $y_p=0.1$, there is only a tiny range in density where it becomes more abundant when $4n$ is included. Looking at $^3$He, this never happens: its abundance is always higher in the calculation with $4n$, in all the temperatures and proton fractions considered. We also observe that the higher the scalar coupling, the higher the abundance, shifting also the dissolution density of the clusters to larger values.

The bottom panels of these figures show the ratio of the chemical equilibrium constants with and without $4n$. For $T=10$ and 20 MeV, the neutron-rich clusters, such as $^6$He, show an increase when the tetraneutron is included in the calculation, in the high-density limit. For the lowest temperature considered, this is also seen, though in a smaller case, and the effect being more predominant in the low-proton fraction case.

We now look in more detail at the mass fractions of the proton and neutron background gas  and of  the $^4$He fractions, particles that have a special role on the transport coefficients (the first two) or on the dissolution  of the accretion disk of a neutron star binary merger \citep{Rosswog2015}. In order to allow for a more clearer discussion, they have been represented in Fig. \ref{fig4}, where they are compared for the three temperatures considered in the present study (4, 10 and 20 MeV) and the two proton fractions (0.3 and 0.1). Some conclusions are in order: i) the effect of the inclusion of the $4n$ is higher in the low-temperature and low-proton fraction case as already discussed before; ii) in a rich neutron environment the effect is quite dramatic with a possible increase of the proton fraction of more than 500\% and the $\alpha$-particle fraction of $\sim 300\%$  if $T=4$ MeV and $x_{sj}=0.92$ are considered (notice that we are plotting the ratios divided by two in this panel). Even for $T=10$ MeV, the increase of the proton fraction is about 100\% and of the $\alpha$ fraction $\sim 33\%$; iii) the neutron fraction is directly affected suffering a noticeable reduction, although not so strong as the proton fraction. It is observed that the reduction is $\lesssim 0.5$ and does not vary much with temperature.

\subsection{The chemical equilibrium constants}

\begin{figure}
 \begin{tabular}{c}
 \includegraphics[width=0.45\textwidth]{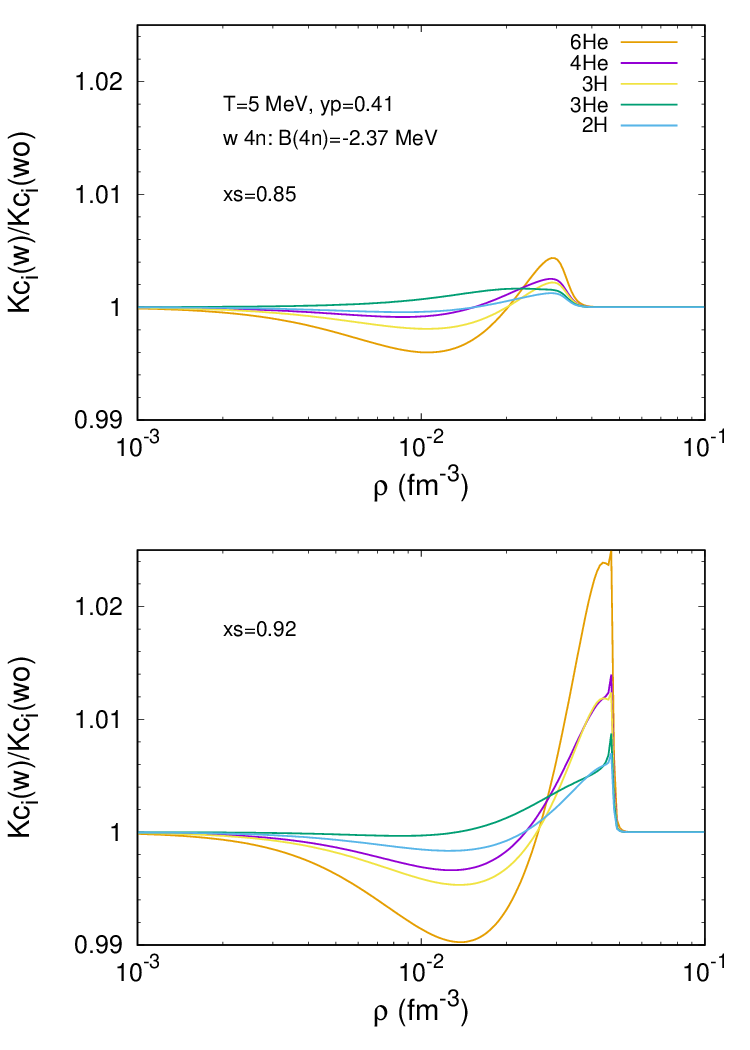} 
  \end{tabular}
  \caption{The ratio between the chemical equilibrium constants with and without the tetraneutron, $K_c[i]$, as a function of the density. In the calculation with the $4n$, we take its binding energy as $B^0_{4n}=-2.37$ MeV. The FSU EoS is considered, and the scalar cluster-meson coupling is fixed to $x_s=0.85$ (top) and $x_s=0.92$ (bottom). We fix the proton fraction to $y_p=0.41$ and the temperature to 5 MeV. } 
\label{fig5}
\end{figure}

Since these clusters are also observed in heavy-ion collisions, we plot in top panel of Fig.~\ref{fig5} the ratio of the chemical equilibrium constants of the clusters in a calculation with and without the tetraneutron, considering $T=5$ MeV and a proton fraction of 0.41, which is the value found in the Texas experiment \citep{Qin2012}, and the scalar cluster-meson coupling fixed to 0.85. We have considered one of the temperatures measured that is more sensitive to the presence of tetraneutrons. As discussed before, we see that the difference in the chemical equilibrium constants with and without the $4n$ is not large, except for the high-density limit where we see that the most neutron-rich cluster increases. Notice that the experimental data were measured for a quite symmetric system, $y_p=0.41$, and the  tetraneutron plays an important role in very asymmetric matter.

In the bottom panel, we use for  the scalar cluster-meson coupling $x_{sj}=0.92$, as determined in \cite{PaisPRL}. As  discussed before, the effect of the tetraneutron inclusion is stronger and this is reflected on the  chemical equilibrium ratio  which increases for a larger coupling.

\section{Conclusions}

We have analyzed the effects of the presence of the $4n$ cluster in hot matter formed in astrophysical environments such as binary NS mergers or core-collapse supernova matter. In particular, how their presence will affect the abundances of the other light clusters was thoroughly discussed. 

This discussion was undertaken within the framework of RMF description of nonhomegenous matter calibrated to  nuclear matter properties \citep{Pais18} and the formation of light clusters in heavy ion collisions \citep{Qin2012,PaisPRL}. This framework allows the inclusion of light clusters as new degrees of freedom and the couplings of the cluster fields to the mesons inherent to the model allow to modulate the effects of the medium, in particular, their dissolution.

It was shown that the presence of $4n$ increases the abundances of free protons as well as of $\alpha$ particles while decreasing the abundance of free neutrons. The effects are stronger in very neutron-rich matter and for lower temperatures.  For $T=5$~MeV, the abundance of free protons could increase by a factor of 5 at the peak of the clusters fractions that occurs at $\sim 0.02$ fm${^-3}$, having, therefore, an effect on the transport properties, such as the conductivity. Also an increase of the $\alpha$ fractions may affect the evolution of the dissolution of the disk formed in a NS merger as discussed in \cite{Rosswog2015}. 

Presently, HIC occur for quite symmetric systems and, therefore, the presence of $4n$ will be difficult to detect. In the best conditions, i.e. lowest temperature detected, we did not find an effect above 2\%, which is certainly difficult to detect.

\section*{Acknowledgements}
H.P. thanks J. Natowitz and G. R\"opke for useful discussions and for bringing to her attention the work of \cite{Duer22}. H.P. and C.P. acknowledge the FCT (Portugal) Projects No. 2021.09262.CBM, 2022.06460.PTDC and UID/FIS/04564/2020. C.A. and M.A.P.G. acknowledge the support of the Agencia Estatal de Investigaci\' on through the grants PID2019-107778GB-100, PID2022-137887NB-I00, and from Junta de Castilla y Le\'on, SA096P20 project.

\appendix
\section{The Cauchy Distribution}\label{append}

The Cauchy (or Lorentz, or Breit-Wigner) probability density is defined as (see e.g. \cite{Ivanytskyi19,Andronic05,Cauchy08})
\begin{align}
f(x;x_0,\gamma)=\frac{1}{\pi\gamma(1+(\frac{x-x_0}{\gamma})^2}
\end{align}
All the moments of this distribution are undefined, since
\begin{align}
M_p=\int_{-\infty}^{\infty}dx\ x^{p} f(x;x_0,\gamma)
\end{align}
diverges. In particular, expected value, variance and curtosis are undefined. This distribution is symmetric and its median is $x_0$, which is the central value. The $x_0-\gamma$ and $x_0+\gamma$ points are the first and third quartiles respectively.

The distribution function of the Cauchy probability density is 
\begin{align}
&F_{{\rm Cauchy}}(x) = \nonumber\\
&\int_{-\infty}^{x} dx \frac{1}{\pi\gamma(1+(\frac{x-x_0}{\gamma})^2)} = \frac{1}{\pi}\arctan\left(\frac{x-x_0}{\gamma}\right) +\frac12,
\end{align}
which is the integrated probability that the random variable $X_{\rm Cauchy}$ satisfies $X_{\rm Cauchy}<x$

In the normal distribution, the variance $\sigma$ of the distribution, which is identified with the uncertainty satisfies that
\begin{align}
P(x_0-\sigma<X_{Normal}<x_0+\sigma) \approx 0.68,
\end{align}
being $x_0$ the mean or expected value of the normal distribution.

As we are dealing with a distribution of mass under the Cauchy law, so to establish an analogy with the Normal distribution we propose to identify
the median with the mean. To build an error interval with a 68\% probability as in the Normal distribution we need to calculate the two points such that
\begin{align}
&P(x_1<X_{\rm Cauchy}<x_2) = \int_{x_1}^{x_2} dx f(x;x_0,\gamma) = 0.68 \nonumber \\
&= F_{\rm Cauchy}(x_2) -F_{\rm Cauchy}(x_1).
\end{align}
As the probability density is symmetric
\begin{align}
&F_{\rm Cauchy}(x_2) = 0.84\nonumber\\
&F_{\rm Cauchy}(x_1) = 0.16
\end{align}
and $x_{1,2}$ will satisfy
\begin{align}
&x_1 = x_0 + R\nonumber \\
&x_1 = x_0 - R.
\end{align}

So as
\begin{align}
\frac{1}{\pi}\arctan\left(\frac{x_1-x_0}{\gamma}\right) +\frac12 = 0.16,
\end{align}
then, straightforwardly,
\begin{align}
x_1&=x_0+\gamma\tan{(-0.34\pi)}=x_0-\gamma\tan{(0.34\pi)}\nonumber \\
&\approx x_0-1.818993\gamma
\end{align}
and subsequently,
\begin{align}
x_2=x_0+\gamma\tan{(0.34\pi)}\approx x_0+1.818993\gamma.
\end{align}

\end{document}